# Combined frequency-amplitude nonlinear modulation: theory and applications

Giancarlo Consolo[1,2,3], Vito Puliafito[3], Giovanni Finocchio[3], Luis Lopez-Diaz[4], Roberto Zivieri[1,2], Loris Giovannini[1,2], Fabrizio Nizzoli[1,2], Giovanna Valenti[5], Bruno Azzerboni[3]

[1] CNISM, Research unit of Ferrara, Italy
[2] Department of Physics, University of Ferrara, 44100 Ferrara, Italy
[3] Department of Matter Physics and Electronic Engineering, University of Messina, 98166 Messina, Italy
[4] Department of Applied Physics, University of Salamanca, 37008 Salamanca, Spain
[5] Department of Sciences for Engineering and Architecture, University of Messina, 98166 Messina, Italy

*Abstract* — **In this work we formulate a generalized theoretical model to describe the nonlinear dynamics observed in combined frequency-amplitude modulators whose characteristic parameters exhibit a nonlinear dependence on the input modulating signal. The derived analytical solution may give a satisfactory explanation of recent laboratory observations on magnetic spin-transfer oscillators and fully agrees with results of micromagnetic calculations. Since the theory has been developed independently of the mechanism causing the nonlinearities, it may encompass the description of modulation processes of any physical nature, a promising feature for potential applications in the field of communication systems.**

## I. INTRODUCTION

IN communication systems, modulation is the process of varying the characteristic parameters of a periodic high-frequency wave ("carrier"), in accordance with a low-frequency information signal ("modulating"), to obtain a "modulated" signal. The carrier wave is generally a sinusoidal waveform characterized by three modifiable parameters, amplitude, frequency and phase, and the corresponding analog modulation processes are referred to as amplitude modulation (AM), frequency modulation (FM) and phase modulation (PM), respectively [1].

In a classical *linear* FM process, the instantaneous frequency of the output signal undergoes a time variation proportional to the information contained in the input modulating signal. In reality, in addition to the basic type of modulation, other spurious analog modulation processes can take place simultaneously distorting the former [1], [2].

There exist, besides, some real cases in which the input-output characteristic is intrinsically nonlinear [3]-[7] and the resulting behavior would be that of a *nonlinear* modulator.

Such a mechanism is observed *e.g.* in magnetic spin-transfer oscillators [8]-[18], where the simultaneous action of a *dc* and an *ac* spin-polarized current may excite a persistent magnetization dynamics which was classified as a "pure" nonlinear FM (NFM) process [9]. In the pioneering experiment of [9], two intriguing features were reported: (i) the frequency of the carrier wave shifts with the increase of the amplitude of the modulating signal; (ii) the spectrum contains sidebands that, even though symmetrically located with respect to the carrier frequency, present different amplitudes. The attempt to justify these dynamics on a physical basis was carried out by analytical models and numerical macrospin calculations, but without fully succeeding in this goal [9].

Here we develop a more sophisticated analytical model mainly based on the idea that the nonlinear dependence of both frequency and amplitude of the modulated signal on the information-carrying input signal has to be included [6]. Our approach represents thus a generalized model of "combined" nonlinear FM-AM (NFAM) modulation. The proposed approach has also the characteristics of an universal model, as it does not depend on the physical phenomenon which gives rise to the above-cited nonlinearities. In fact, since these nonlinear dependences are included in the model as Taylor polynomial expansions (of a given order) around a bias point, the model may encompass whatever nonlinear functional dependence. In other words, in spite of the validation of our theoretical model will be carried out for the case of a spin-torque nano-oscillator, our approach might find application in other research fields as well, as it will be mentioned later on in the paper.

In order to test the validity of the proposed analytical model, we compare the theoretical results with those obtained numerically by means of micromagnetic simulations. It is worth noticing that the resulting agreement does not require any adjusting parameter into the fitting procedure.

Corresponding author: G. Consolo (e-mail: consolo@ingegneria.unime.it).

The paper is organized as follows.

Section II describes the two analytical models of nonlinear modulation. In particular, we discuss first the NFM model, which represents the initial approach used to investigate the dynamics of spintronic modulators [9]. After that, we present our generalized NFAM model.

Section III presents details on the micromagnetic framework used to carry out the numerical experiments on the test case of a spintronic frequency modulator.

Sections IV and V are devoted to the discussion of results and conclusions, respectively.

## II. NONLINEAR MODELS OF ANALOG MODULATION

Let us denote by $c(t) = A_c \cos(2\pi f_c t)$ the temporal evolution of the *carrier signal* having amplitude $A_c$ and frequency $f_c$. As $c(t)$ is a pure sinusoidal wave, it brings no information. On the contrary, we refer to $m(t)$ as the baseband *modulating signal* which carries specifications on the message to be transmitted. Ideally, a FM signal is in the form:

$$s(t) = A_c \cos\left[\theta_i(t)\right], \qquad (1)$$

where $A_c$ is the constant carrier amplitude and $\theta_i(t)$ is the instantaneous phase of the modulated signal, related to the instantaneous frequency by $f_i(t) = (1/2\pi) d\theta_i(t)/dt$. In a typical FM system [1], the instantaneous frequency varies *linearly* with the modulating signal: $f_i(t) = f_c + k_f m(t)$ ($k_f$ is called frequency sensitivity). According to the previous definitions, eq.(1) may be re-written as:

$$s(t) = A_c \cos\left[2\pi f_c t + 2\pi k_f \int_0^t m(\tau) d\tau\right]. \qquad (2)$$

Since through the whole paper we are mainly interested in the derivation of the Fourier spectra of the modulated signals, the initial phase $\theta_i(0)$ in eq.(2) has been neglected. This constant term does not bring indeed any contribution to our purposes.

The modulated signal defined by eq.(2) is a nonlinear function of the modulating signal $m(t)$, which makes FM an intrinsic *nonlinear* modulation process. Nevertheless, because of the linearity between $f_i(t)$ and $m(t)$, we refer to this phenomenon as a *linear* modulation process.

In the following, we shall refer to a different situation where the instantaneous frequency depends *nonlinearly* on the modulating signal, such as:

$$f_i(t) = \sum_{h=0}^{v} k_h m^h(t). \qquad (3)$$

According to this formulation, $k_0 = f_c$ and $k_h$ (for $h > 1$) are $h$th-order frequency sensitivity coefficients.

We also investigate on the possibility that a similar nonlinear dependence may affect the signal amplitude. In this case, the dependence of the output amplitude on the input signal may be included by generalizing the formulation of the term $A_c$ (defined in eq.(1)) as follows:

$$A_c(t) = \sum_{k=0}^{u} \lambda_k m^k(t) \qquad (4)$$

where $\lambda_k$ identify $k$th-order amplitude sensitivity coefficients.

In the field of communications systems, the simultaneous validity of eqs.(3) and (4) may be understood by assuming that AM and FM phenomena take place at the same time. From a more general viewpoint, these assumptions simply describe a non-negligible intrinsic nonlinear dependence of both amplitude and frequency on the input stimulus, as it happens in many real systems (mechanical [3]-[5], magneto-electronic [6-18] and optical [19], to cite a few). However, these functional dependences might be quite complicated, so that the derivation of the corresponding Fourier spectra might be, in some cases, precluded. The usage of polynomial series to express those nonlinear relationships overcomes this difficulty.

With this is mind, in the following we discuss two theoretical models of nonlinear modulation:
(a) "NFM", which deals with a "pure" FM process where the only hypothesis of nonlinearity made in eq.(3) is included [9];
(b) "NFAM", which deals with the new formulation of a "combined" FM-AM process where both the hypotheses of nonlinearity made in eqs.(3) and (4) are included.

In order to illustrate the properties of both models we shall consider, for simplicity, a single-tone modulation in which the modulating signal is in the form of a sinusoidal wave having amplitude $A_m$ and frequency $f_m$: $m(t) = A_m \cos(2\pi f_m t)$.

We focus first on the "NFM" model, and assume the relationship $f_i(t)$ versus $m(t)$ to be represented as a Taylor expansion around $f_c$, as in eq.(3), and the carrier amplitude $A_c$ to be constant. Substituting the expression of $m(t)$ into (3), and (3) into (1), after some algebraic steps one ends up with the expression of a *nonlinear* FM output signal:

$$s(t) = A_c \cdot \cos\left[\omega_c^i t + \sum_{h=0}^{v} \beta_h \sin(h\omega_m t)\right] \qquad (5)$$

where $v$ is the order of the polynomial (3), $\omega_m = 2\pi f_m$ is the angular frequency of the modulating signal, $\beta_0 = 0$ and $\beta_h$ (for $h > 1$) represent $h^{th}$-order frequency modulation indexes given by a linear combination of the coefficients $k_h$, and

$$f_c^I = \frac{\omega_c^I}{2\pi} = \left( f_c + \sum_{h=0}^{\nu} \eta_h (k_h) A_m^{\ h} \right) \qquad (6)$$

represents the central frequency of the modulated signal. In (6), the coefficients $\eta_h (k_h)$ obtained for odd $h$ (and $h = 0$) are identically zero, so that the central frequency $f_c^I$ is shifted, with respect to the frequency of the un-modulated carrier $f_c$, by an amount which depends on the amplitude of the modulating signal $A_m$ through the coefficients with even index $h$ only. Such a property is the *first* indication of a nonlinear modulation process. In fact, differently from the case of linear modulation [1], the central frequency can here undergo either a blue-frequency shift ($f_c^I$ increases with increasing $A_m$) or a red-frequency shift ($f_c^I$ decreases with increasing $A_m$) according to the signs of the even coefficients of the Taylor expansion (3).

By using the formalism of complex envelopes [1], the Fourier transform of eq.(5) can be approximated as follows:

$$S(f) = \frac{A_c}{2} \sum_{\substack{\zeta_p = -\infty \\ p = 0, \dots, \nu}}^{+\infty} Re\left\{ \prod_{h=0}^{\nu} \mathcal{J}_{\zeta_h} (\beta_h) \right\} \cdot$$
$$\cdot \left[ \delta\left( f - f_c^I - f_m \sum_{h=0}^{\nu} h\zeta_h \right) + \delta\left( f + f_c^I + f_m \sum_{h=0}^{\nu} h\zeta_h \right) \right] \qquad (7)$$

where $\mathcal{J}_{\zeta_h} (\beta_h)$ is the $\zeta_h^{th}$-order Bessel function of the first kind and argument $\beta_h$ (by using this notation $\zeta_0 = 0$). In this equation, the arguments $\beta_h$ might be either positive or negative according to the signs of the coefficients of the Taylor series (see eq.(3)). For negative $\beta_h$, the corresponding Bessel function can assume a complex value. However, the modulus of the imaginary part is several orders of magnitude smaller than the one of the real part and can be safely neglected.

According to the "NFM" model (eq.(7)), the frequency spectrum of a nonlinear FM signal consists of a central frequency $f_c^I$, shifted with respect to the frequency of the un-modulated carrier, and an infinite number of sidebands symmetrically located at $f_c^I \pm l f_m$ ($l$ is the positive integer identifying the sideband order). In this case, however, upper and lower sidebands of $l^{th}$-order (obtained from (7) under the constrain $\sum_{h=0}^{\nu} h\zeta_h = \pm l$) generally exhibit different amplitudes owing to their dependence on the product of Bessel functions of different order. The existence of sidebands symmetrically located with respect to the shifted carrier but asymmetric in amplitude represents the *second* main difference between nonlinear [9] and linear modulation [1].

In order to introduce the "NFAM" model, we generalize, according to the hypothesis made in eq.(4), the formulation given in (1) by expressing a combined FM-AM signal in the time domain as follows:

$$s(t) = A_c (t) \cos\left[ \theta_i (t) \right] \qquad (8)$$

By combining eqs.(5) and (4) within eq.(8) and performing the Fourier transform of that function, we derive the Fourier spectrum of a nonlinear combined AM-FM signal:

$$S(f) = \frac{1}{4} \sum_{k=0}^{u} \gamma_k \sum_{\substack{\zeta_j = -\infty \\ j = 0, \dots, \nu}}^{+\infty} Re\left\{ \prod_{i=0}^{\nu} \mathcal{J}_{\zeta_i} (\beta_i) \right\} \cdot$$
$$\cdot \left\{ \delta\left[ f - f_c^I - \left( \sum_{i=0}^{\nu} i\zeta_i + k \right) f_m \right] + \delta\left[ f + f_c^I + \left( \sum_{i=0}^{\nu} i\zeta_i + k \right) f_m \right] + \right.$$
$$\left. \delta\left[ f - f_c^I - \left( \sum_{i=0}^{\nu} i\zeta_i - k \right) f_m \right] + \delta\left[ f + f_c^I + \left( \sum_{i=0}^{\nu} i\zeta_i - k \right) f_m \right] \right\} \qquad (9)$$

where $u$ is the order of the polynomial (4) and the amplitude modulation indexes $\gamma_k$ are linear combinations of the coefficients $\lambda_k$ of that Taylor series. Eq.(9), which represents the new formulation "NFAM" proposed in the present work, determines the structure of the frequency spectrum of a combined FM-AM signal having whatever nonlinear dependence in both frequency (3) and amplitude (4) as a function of the modulating signal. Formulation (9) is qualitatively similar to (7) as it predicts the same frequency shift of the carrier signal ($f_c - f_c^I$) together with the existence of symmetric sidebands having different amplitudes. Despite of this, it contains a remarkable difference with respect to (7). In fact, depending on the values of the coefficients $\gamma_k$, the two models might provide substantially different values for the amplitude of sidebands.

## III. MICROMAGNETIC MODEL

In order to estimate whether the difference between eqs.(7) and (9) is really quantitative or, conversely, it brings just a qualitative contribution, we test the theoretical models described in the previous section by taking into account the nonlinear modulation dynamics observed in magnetic nanocontact devices [6], [8]-[16] subjected to the simultaneous action of a *dc* and an *ac* spin-polarized current [9]. As introduced previously, we compare the theoretical results with those obtained by means of micromagnetic simulations.

A magnetic nanocontact device consists of a layered structure made by two extended magnetic layers (a thicker

"fixed" or "pinned" layer (PL) and a thinner "free" layer (FL)) separated by a nonmagnetic spacer. A metallic circular contact of radius $R_c$ is lithographically defined on the top of the FL, providing the opportunity to apply a perpendicular-to-plane current in a reduced region of the FL only (see Fig.1).

## FIG.1 HERE

The dynamics of the magnetization vector of the FL in both time and spatial domain, $\mathbf{M} = \mathbf{M}(t, \mathbf{r})$, is governed by the Landau-Lifshitz-Gilbert-Slonczewski (LLGS) equation [6]-[8]:

$$\frac{\partial \mathbf{M}}{\partial t} = \gamma \left[ \mathbf{H}_{\text{eff}} \times \mathbf{M} \right] + \frac{\alpha}{M_0} \left[ \mathbf{M} \times \frac{\partial \mathbf{M}}{\partial t} \right] + f\left(r/R_c\right) \frac{\sigma I}{M_0} \left[ \mathbf{M} \times (\mathbf{M} \times \mathbf{p}) \right] \tag{10}$$

where $\gamma$ is the gyromagnetic ratio and $\mathbf{H}_{\text{eff}}$ is the effective magnetic field which includes magnetostatic, exchange, and Zeeman contributions. For simplicity, in our model we neglect the current-induced (Oersted) magnetic field, the magnetostatic coupling between the two ferromagnetic layers and thermal fluctuations as they do not play a significant role in this context. We also ignore the magnetocrystalline anisotropy in the FL, which is an usual assumption for magnetically soft Permalloy layers. The second term in the right-hand side of eq.(10) is the phenomenological magnetic damping torque written in the traditional Gilbert form ($\alpha$ is the damping constant) and $M_0 = |\mathbf{M}|$ is the saturation magnetization of the FL. The last term is the Slonczewski spin-transfer torque that is proportional to the bias current $I$. The function $f\left(r/R_c\right)$ describes the spatial distribution of the current across the area of the nanocontact. In the simplest case of uniform current density distribution, $f\left(r/R_c\right) = 1$ if $r < R_c$ and $f\left(r/R_c\right) = 0$ otherwise. The coefficient $\sigma$ is related to the dimensionless spin polarization efficiency $\varepsilon$ by $\sigma = \varepsilon g \mu_B / 2eM_0 S d_{\text{FL}}$, where $g$ is the spectroscopic Landé factor, $\mu_B$ is the Bohr magneton, $e$ is the absolute value of the electron charge, $d_{\text{FL}}$ is the FL thickness and $S = \pi R_c^2$ is the nanocontact area. The unit vector $\mathbf{p}$ defines the spin-polarization direction which coincides with the equilibrium direction of the PL magnetization. It is obtained by solving Brown's equation for the PL: $\mathbf{p} \times \mathbf{H}_{\text{eff}} = 0$ with no current.

In our approach, the LLGS equation is numerically solved by using our own three-dimensional (3D) finite-difference time-domain (FD-TD) micromagnetic code that employs a fifth-order Runge-Kutta integration scheme [20].

The parameters used to simulate the current-induced spin-wave dynamics in a Permalloy FL are: thickness $d_{\text{FL}} = 5$ nm,

nanocontact radius $R_c = 20$ nm, spin-polarization efficiency $\varepsilon = 0.25$, saturation magnetization $\mu_0 M_0 = 0.7$ T, spectroscopic Landé factor $g = 2.0$, and exchange stiffness constant $A_{FL} = 1.4 \times 10^{-11}$ J/m.

The magnitude $|\mathbf{H}_{\text{ext}}|$ of the external bias magnetic field is chosen to be $\mu_0 \mathbf{H}_{\text{ext}} = 0.8$ T and the field vector $\mathbf{H}_{\text{ext}}$ is directed at 80 degrees out of the structure plane.

The parameters used to compute the equilibrium magnetic state of the Co-based PL are: thickness $d_{\text{PL}} = 20$ nm, saturation magnetization $\mu_0 P_0 = 1.88$ T and exchange stiffness constant $A_{PL} = 2.0 \times 10^{-11}$ J/m.

In our numerical experiments we restrict our study to a limited square computational region as large as $L \times L \times d_{FL} = 800 \times 800 \times 5$ nm$^3$, by using a 2D mesh of discretization cells having sizes $4 \times 4 \times 5$ nm$^3$. As discussed in previous works [12]-[15], to avoid the spurious spin-wave reflection from the computational boundaries we implement abrupt absorbing boundary conditions. Compatibly with a reasonable computational time, each simulation is long enough to assure a spectral accuracy as fine as 0.5 MHz.

Within our micromagnetic approach, the output signal is identified through the time-variations of the Giant MagnetoResistance (GMR) signal [21]. It is assumed to be proportional to the angle of misalignment $\varphi(t)$ between the directions of the local magnetization vectors of the two ferromagnetic layers and, in first approximation, can be expressed as $\tilde{g}_{av}(t) = \left\langle \left(1 - \cos \varphi_k(t)\right)/2 \right\rangle_{\text{contact}}$, where we average the local contributions $\varphi_k(t)$ over the contact area [12]-[15]. Results will demonstrate that such a simplistic assumption is enough accurate for the characterization of the observed nonlinear dynamics.

## IV. RESULTS

The numerical analysis consists of two stages, corresponding to the analysis without and with the modulating signal $m(t)$, respectively.

During the first stage, we carry out the identification of the parameters which appear in the relationships among $f_i(t)$, $A_c(t)$ and $m(t)$ (eqs.(3) and (4) in the manuscript). We consider thus a $dc$ input bias current $I = I_{dc}$ to generate a stable microwave output signal [9], [12]-[15] that we associate to the carrier wave. Since no modulating signal has been considered, such condition implies $\tilde{g}_{av}(t) = c(t)$. Based on our parameter set, we choose the value $I_{dc_0} = 18$ mA as the bias point which corresponds to the excitation of propagating spin-wave having frequency $f_c = 17.725$ GHz and amplitude $\lambda_0 = 0.34341$ (arb. units). In spite of our system exhibits a nonlinear behavior in a large interval of bias currents (we explored the range $13 \text{ mA} < I < 21 \text{ mA}$), we are interested in a small oscillations analysis around that bias point. To this aim, we sweep the current $I$ in a restricted interval of currents under the constrain $\left| I - I_{dc_0} \right| < 1.5 \text{ mA}$ and report the corresponding values of $f_i$ and $A_c$ as in Fig.2. From these values, it is possible to approximate the nonlinear relationships $f_i(I)$ and $A_c(I)$ by computing analytically the Taylor series around the bias point: $f_i\left( I - I_{dc_0} \right)$ and $A_c\left( I - I_{dc_0} \right)$. Since the modulation process [9], simulated during the second stage of the present analysis, is due to the superposition of a carrier wave and a modulating signal, the latter implemented as a sinusoidal $ac$ current $I_{ac} = m(t) = A_m \cos\left( \omega_m t \right)$, the total bias current will be expressed as: $I = I_{dc_0} + m(t)$. Under these circumstances, the modulating signal is given by $m(t) = \left( I - I_{dc_0} \right)$. Moreover, the previous Taylor expansions $f_i\left( m(t) \right)$ and $A_c\left( m(t) \right)$, become explicitly dependent on the modulating signal under the constrain $A_m \ll I_{dc_0}$. This method allows the identification of the $k_h$ and $\lambda_k$ coefficients. As a result of this identification procedure, we get a fourth-order polynomial for the expression $f_i\left( m(t) \right)$ (*i.e.* $v = 4$ in eqs.(7) and (9) of the manuscript) with coefficients: $k_1 = 0.155 \text{ GHz/mA}$, $k_2 = -0.013 \text{ GHz/mA}^2$, $k_3 = 0.00883 \text{ GHz/mA}^3$, $k_4 = -0.0016 \text{ GHz/mA}^4$ (Fig.2, solid black line), whereas the best fit which represents the function $A_c\left( m(t) \right)$ is expressed in the form of a third-order polynomial ($u = 3$ in eq.(9)), with coefficients: $\lambda_0 = 0.34341$, $\lambda_1 = 0.0535 \text{ mA}^{-1}$, $\lambda_2 = -0.014 \text{ mA}^{-2}$, $\lambda_3 = 0.0007 \text{ mA}^{-3}$ (Fig.2, dashed red curve). The frequency and amplitude modulation indexes, $\gamma_0 = \lambda_0 + \frac{\lambda_2 A_m^2}{2}$, $\gamma_1 = \lambda_1 A_m + \frac{3\lambda_3 A_m^3}{4}$, $\gamma_2 = \frac{\lambda_2 A_m^2}{2}$, $\gamma_3 = \frac{\lambda_3 A_m^3}{4}$, $\beta_0 = 0$, $\beta_1 = \left( k_1 A_m + \frac{3k_3 A_m^3}{4} \right) / f_m$, $\beta_2 = \left( k_2 A_m^2 + k_4 A_m^4 \right) / 4 f_m$, $\beta_3 = k_3 A_m^3 / 12 f_m$, $\beta_4 = k_4 A_m^4 / 32 f_m$, can be obtained by substituting the above-computed coefficients.

**FIG.2 HERE**

During the second stage of our approach, where the modulation process takes place, the output signal corresponds to the modulated signal, $\tilde{g}_{av}(t) = s(t)$, as defined in eqs.(1) and (8). The analysis is performed by considering modulating signals having a fixed frequency $f_m = 500 \text{ MHz}$ and amplitude $A_m$ which varies in the range $0 \div 1.5 \text{ mA}$.

We specifically examine whether the discussed "NFM" and "NFAM" models can *quantitatively* predict both the amount of the frequency shift of the carrier wave and the difference existing in the amplitude of the symmetrically located sidebands (the previously discussed features (i) and (ii), respectively), as function of the amplitude of the modulating signal $A_m$.

We analyze first the relationship between $f_c^I$ and $A_m$ (feature (i)).

The results of the comparison between analytical and numerical results are shown in Fig.3(a). It should be noticed that no differences are theoretically expected, and numerically detected, between results of model "NFM" (7) and "NFAM" (9). Such a conclusion agrees with the theoretical prediction that the amplitude dependence (4) (*i.e.* the additive AM modulation process) does not shift the central frequency (6). We also notice that, since the even coefficients $k_h$ are both negative, the central frequency $f_c^I$ decreases (red-shifts) with increasing the amplitude $A_m$.

**FIG.3 HERE**

We analyze now the relationship between the amplitude of $l$th-order sidebands $\left| S\left( f = f_c^I \pm l f_m \right) \right|$ and $A_m$ (feature (ii)).

For simplicity, we limit our analysis to the study of first ($l = 1$) and second ($l = 2$) order sidebands. High-order ($l > 2$) sidebands are characterized by amplitude values which are several orders of magnitude smaller than the low-order ($l \leq 2$) ones, so that their contribution can be safely neglected. To summarize the investigated relationship in only one curve, we introduce a dimensionless variable $\Psi_l$ which represents the ratio between the amplitude of upper and lower sideband of $l$th-order:

$$\Psi_l = \left| S\left( f = f_c^I + l f_m \right) \right| / \left| S\left( f = f_c^I - l f_m \right) \right|. \quad (11)$$

If the modulated signal obtained by means of the described procedure were the result of a "pure" nonlinear FM process, as done in [9], a substantial agreement between results of numerical calculations and those derived from the theoretical "NFM" model would be achieved. Results reveal, on the contrary, a remarkable disagreement between them (in Fig.4, compare symbols (numerical) with dotted lines (theoretical)). This led us to investigate the origin of such discrepancy.

**FIG.4 HERE**

First, from the numerical point of view, the output amplitude almost doubles its value when passing from $I = 16.5 \text{ mA}$ to $I = 19.5 \text{ mA}$, as observed in Fig.2 (dashed curve). It is a clear indication that the assumption $A_c$ = constant is inappropriate for this context and must be discarded. This is also the main reason why the approach carried out in [9] could not accurately reproduce the different sidebands amplitudes. On the other hand, it should be pointed out that time-domain analyses of the microwave signals corresponding to nanoscale dynamics have been carried out only recently [22,23], so that they were not available at the date of publication of Ref.9. This is why micromagnetic simulations become essential to gain a qualitative (and often quantitative) understanding of the phenomena which occur, in both time and spatial domain, in the corresponding experiments. The powerful tool of numerical calculations reveals indeed that the modulated signals present some characteristic features (mainly related to the envelope) ascribable to the simultaneous occurrence of an AM process (see Fig.3(b ,c)).

Second, from the theoretical point of view, a theory of microwave generation in current-driven magnetic oscillators demonstrated that the nonlinear shift of the generated frequency with the input current is strictly related to the variation of the projection of the magnetization vector on the precession axis which in turn demands for a change of the precession angle [6]. In fact, any variation of the precession angle reflects, through the GMR effect [21], in a change of amplitude of the output signal. In other words, in magnetic spin-transfer oscillators, the relationship existing between the nonlinear frequency shift and the output amplitude cannot be disregarded [8], [24], [25].

All the above reasons suggested us to account for an additive nonlinear dependence of the output amplitude on the input bias current, as in eq.(4), and to assume that a combined AM-FM process take place.

The results of the comparison between numerical calculations and theoretical ones (obtained by using eq.(9)) are shown in Fig.4 (solid lines). The remarkable quantitative agreement confirms therefore the correctness of our conjecture about the simultaneous occurrence of two modulation processes stimulated by a sole input information-carrying signal.

## V. Conclusions

The above presented comparison of the analytical results arising from the generalized nonlinear model of a combined NFAM modulation process with results of micromagnetic calculations performed on a magnetic nanocontact spin-transfer oscillator demonstrates that the "NFAM" model (9) can give an accurate description of the complicate dynamics reported experimentally in [9].

In particular, it can quantitatively describe, without using any adjustable parameter, the two main characteristics of the frequency spectrum reported for a current-driven spintronic modulator: (i) shift of the central frequency and (ii) symmetrically located sidebands with substantial different amplitude.

The key advantage of the "NFAM" model to capture the laboratory dynamics [9] lies in the capability to include, at the same time, the nonlinear dependences of both instantaneous frequency and amplitude on the input modulating signal. This latter contribution, while bringing a null role for the shift of the central frequency (6), becomes fundamental for the exact estimation of the sidebands amplitude.

From a more general point of view, by means of a proper identification of the parameters appearing in eq.(9), the model can describe the behavior of a generic nonlinear analog modulator, independently of both the specific functional dependence of the characteristic parameters on the modulating signal and the mechanism originating the nonlinearity.

The proposed approach might find application also in other research fields, such as sonar communications, where an arbitrary chirp is modeled through a nonlinear combination of an AM and a PM signal (see [26] and references thereinafter). In those cases, both the amplitude and the phase of the modulated signal exhibits a very elaborated nonlinear dependence of time (see, *e.g.*, eqs.(4) and (8) of [26]) which generally prevents the derivation of an analytical solution for the corresponding integrals. By using our method, it would be possible, in principle, to overcome this difficulty and predict the composition of the frequency spectrum of the modulated signal.

Finally, we would like to mention that, after submitting our paper, a recent experiment of analog modulation on spin-transfer oscillators reinforced the validity of our approach [27].

## Acknowledgments

This research was supported by the Spanish government (project MAT2008-04706/NAN), by the Junta de Castilla y Leon (project SA025A08), and by CNISM through "Progetto Innesco".

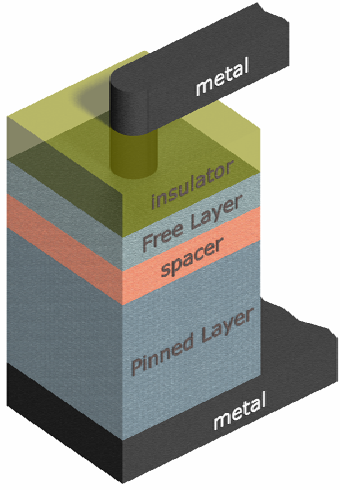

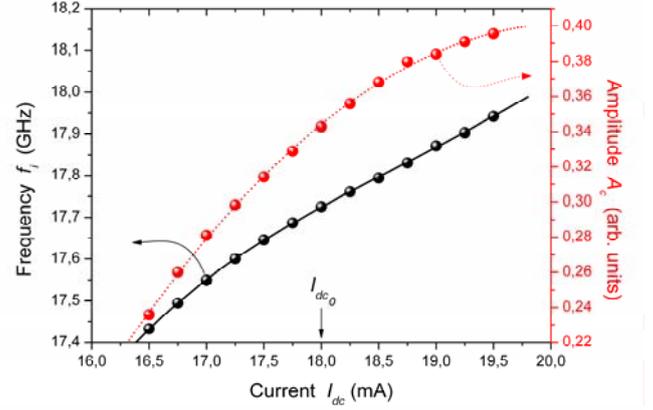

Fig. 2. (Color online) Dependence of frequency $f_i$ (black symbols) and amplitude $A_c$ (red symbols) on the input current $I_{dc}$ as a result of the parameters identification procedure. The corresponding polynomial best-fits are shown by the solid black curve (eq.(3), with $v = 4$ ) and the dotted red curve (eq.(4), with $u = 3$ ), respectively.

Fig. 1. (Color online) Schematics of a magnetic multilayer nanocontact device.

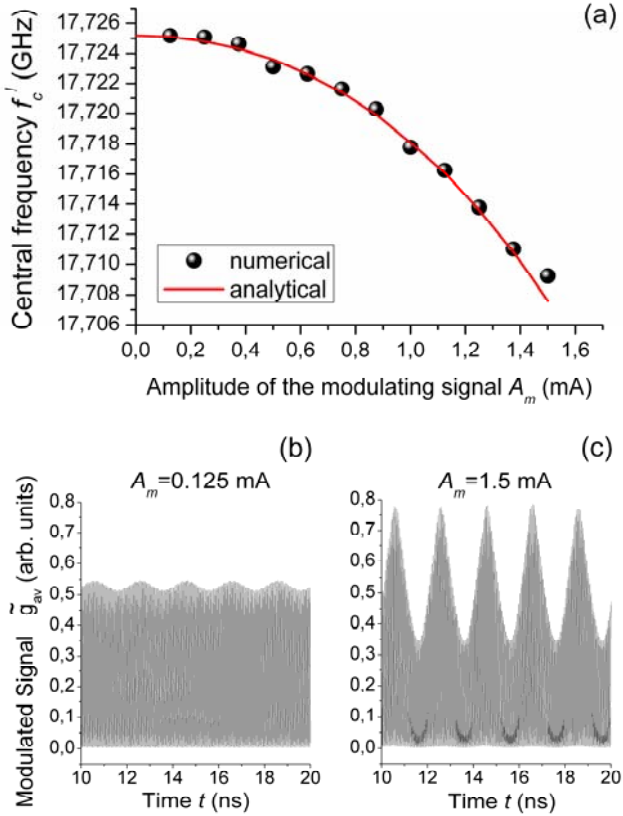

Fig. 3. (a) Dependence of the central frequency $f_c^I$ on the amplitude of the modulating signal $A_m$. Symbols are representative of numerical results whereas the solid line shows the theoretical dependence predicted by both "NFM" and "NFAM" model. In (b) and (c) we show details of the temporal evolution of the modulated signal $s(t)$ for $A_m = 0.125$ mA and $A_m = 1.5$ mA , respectively.

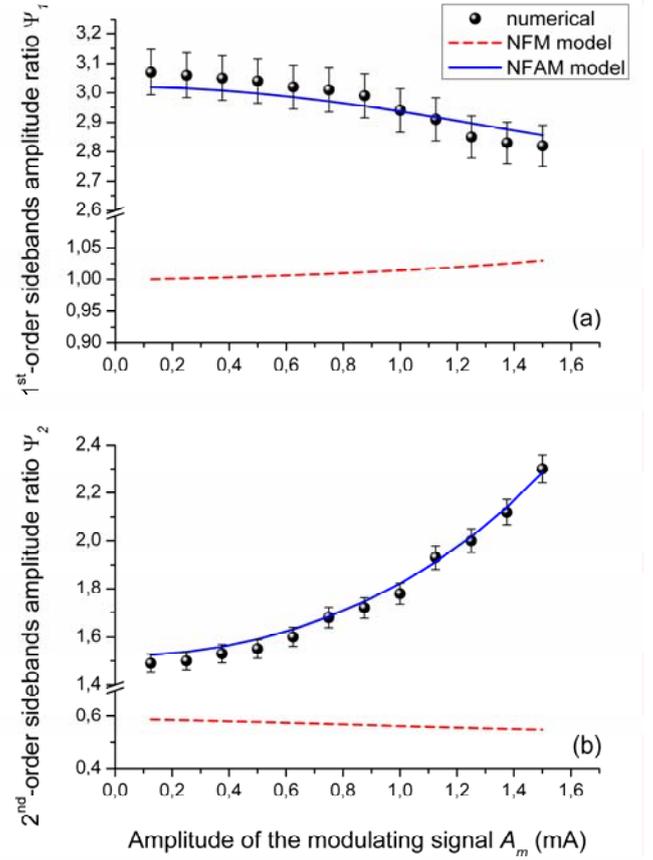

Fig. 4. (Color online) Dependence of $1^{st}$-order (a) and $2^{nd}$-order (b) sidebands amplitude ratio on the amplitude of the modulating signal $A_m$. Symbols, together with error bars, are representative of numerical results, whereas the analytical ones are denoted by lines (dotted lines for "NFM" and solid lines for "NFAM"). Error bars associated to numerical results are representative of a computational error of about 2.5%.